\documentclass[%
 aip,
rsi,%
 amsmath,amssymb,
 reprint,%
]{revtex4-1}

\usepackage{graphicx}
\usepackage{dcolumn}
\usepackage{bm}
\usepackage{nicefrac}

\begin{document}

\preprint{AIP/123-QED}

\title[Rodr\'{i}guez et al.]{RF current condensation in magnetic islands and associated hysteresis phenomena }

\author{E. Rodr\'{i}guez}
 \altaffiliation[Also at ]{eduardor@princeton.edu}

\affiliation{ 
Princeton University, Princeton, NJ, 08540
}

\author{A. H. Reiman}%
 \email{areiman@pppl.gov}
\author{N. J. Fisch}
\affiliation{%
Princeton Plasma Physics Laboratory
}%

\date{\today}

\begin{abstract}
The nonlinear RF current condensation effect suggests that magnetic islands might be well controlled with broader deposition profiles than previously thought possible.  To assess this possibility, a simplified energy deposition model in a symmetrised 1D slab geometry is constructed.  By limiting the RF wave power that can be absorbed through damping, this model describes also the predicted hysteresis phenomena.  Compared to the linear model, the nonlinear effects lead to larger temperature variations, narrower deposition widths,  and more robust island stabilisation.  Although, in certain regimes, the island centre can be disadvantageously shaded because of the nonlinear effects, in general, the RF condensation effect can take place, with current preferentially generated, advantageously, close to the island centre.

\end{abstract}

\maketitle

\section{\label{sec:intro} Introduction:}
Magnetic confinement approaches to fusion rely on the ordered topology of nested magnetic surfaces to prevent the plasma from escaping. Devices such as tokamaks and stellerators are designed towards this; however, in reality, magnetic fields are not perfect and are subject to error fields. These change the magnetic topology and result in magnetic islands appearing at rational surfaces. \par
Magnetic islands are characterised by flat density and temperature profiles due to enhanced transport through them. This reduction of pressure gradients suppresses bootstrap current within the island, generally making the island grow until saturation\cite{Chang,LaHaye97,ZohmEPS97,Gates97}. As a result, the confinement ability of the system decreases, paired with the occurrence of so called neoclassical tearing modes (NTMs)\cite{Buttery_2000}. \par
NTMs were recognised as a source of major disruptions in experiments such as JET\cite{devries,devries14}, and thus their stabilisation  is central. Amongst proposed stabilisation approaches, driving current\cite{reiman,Yoshioka} at the islands with RF waves has stimulated a long list of added efforts,\cite{Sauter2004,Kamendje2005,LaHaye2006,Lahaye08,Henderson08,Lazzari,Volpe,Sauter2010,Bertelli2011,Hennen2012,Smolyakov_2013,Ayten,Borgogno,Volpe2015,Fevrier2016,Wang2015,JC_Li_2017,
Grasso_JPP2016,Grasso18,Poli15}
including many experimental demonstrations\cite{Bernabei,Warrick2000,Gantenbein,Zohm2001,Isayama2000,LaHaye2002,Petty04}. By driving current at the centre of the island using ECCD\cite{fisch80,karney81} or LHCD\cite{fisch78,karney79} one may balance the lack of bootstrap current, and prevent the island growth.\cite{Lazzari} This technique is, however, limited to its application to islands of smaller size due to available power constraints. This makes driving current precisely at their centre difficult, as the deposition width is comparable to the island size. \par
It has been recently suggested\cite{Reiman18} that some of these stringent requirements may be relaxed due to the so called "RF current condensation effect". This effect takes into consideration the non-linear feedback of temperature variations\cite{Westerhof07} resultant from RF wave heating onto the deposition itself. It was found that condensation could improve mitigation as well as reduce radial sensitivity.  
 \par
Formally, in Ref.~[\onlinecite{Reiman18}] this non-linear feedback was modeled using a simplified diffusion energy balance equation that included resonant power deposition for a prescribed profile. Yet, the lack of a dissipation mechanism (eg. radiation) or the unlimited absorbable power from RF waves, gave, as they observed, an nonphysical temperature blow-up beyond a bifurcation point. Here that model is extended to include the damping of the RF wave.  \par
In what follows, this extended model is first introduced in detail. Next, the equation is analytically and numerically solved and a hysteresis effect related to the island heating is described. Having introduced this phenomena, the effects of the non-linearities on different RF deposition schemes are explored. In order to evaluate these effects fairly, comparisons are made to the analytic linear solution, which is taken as representative of current approaches.

\section{Fundamental theoretical model}

The model describes both the temperature variations of a magnetic island and the RF wave deposition. The latter may be described as a wave that is being damped along a ray trajectory, so as to provide the plasma with energy. The temperature of the plasma, driven by the RF waves, is described as part of an energy balance model with thermal diffusion. To construct such a model, various approximations are introduced. \par
 Take, as a starting point, the transport equation\cite{Braginski}, representing the second moment of the Boltzmann equation, to describe temperature, $T$, dynamics. Under the assumption of no significant flows, and considering fast equilibration between electrons and ions, a single equation may be written combining the two-fluid Braginskii equations
\begin{equation}
\frac{3}{2}nk_B\partial_t T-\nabla\cdot(\mathbb{\kappa}\nabla T)=P \label{eqn::difEqn}
\end{equation}
where $n$ is the plasma density, $\kappa$ is the generalised heat conductivity tensor, and $P$ a volumetric power deposition which will be later related to the RF power input. \par
It has been stated as an assumption for Eq.~(\ref{eqn::difEqn}) that electrons and ions are effectively equilibrated. Thus, the equation should only apply to those time scales larger than the typical equilibration time $\tau_\mathrm{eq}$; this constitutes the first temporal constraint: $t\gg\tau_\mathrm{eq}$, where $t$ denotes the time scales that the model is suited to describe. 
\par
A second point is related to the use of temperature as a measure of plasma energy. The concept of temperature customarily applies only to thermalised systems, in which the populations in $v$-space are Maxwellian distributed. However, the continuous injection of RF waves distorts the distribution function locally so as to deposit energy resonantly onto a small fraction of faster electrons (eg. for LH waves $v\approx 4.5v_{Te}$ and for EC $v\approx 3v_{Te}$)\cite{fisch87,karney81}. Hence, the plasma is made up of a Maxwellian bulk with a well defined $T$, but also a resonant minority population. If the bulk $T$ is to represent the total internal energy of the plasma, then the energy drawn locally in $v$-space needs to be redistributed quickly. Here a second time ordering is introduced: collisional thermalisation ($\tau_\mathrm{se}$) and isotropisation ($\tau_\perp$) processes must be faster than the time scales of interest ($t\gg\tau_\mathrm{se},\tau_\perp$). In that case non-Maxwellian features that affect a minute fraction of the total population may be ignored to leading order. Kinetic details will still prove important for $P$. \par
As it stands, Eq.~(\ref{eqn::difEqn}) remains a three dimensional problem, but it may be cast into an approximate reduced 1D problem assuming the following. Consider as it happens (see Table \ref{tab::timeScales})\cite{Spitz53,Westerhof07}, transport over a given magnetic flux surface to be much faster than perpendicular to it; that is, $f_\kappa\equiv\kappa_\parallel/\kappa_\perp\gg1$. In that case, magnetic surfaces will be approximately isothermal, simplifying the derivatives in the diffusive term (i.e. the second term in Eq.~(\ref{eqn::difEqn})). For analytic simplicity, the geometrical factor associated with the particular shape of island flux surfaces will be ignored, using instead a single slab coordinate $x$ (one may think of making a cut to an elongated, narrow island). This reduced form is,
\begin{equation}
\frac{3}{2}nk_B\partial_t T-\partial_x(\mathbb{\kappa_\perp}\partial_x T)=P \label{eqn::dif2}
\end{equation}
As a result of this slab adaptation, areal weighting is made equal for all points, though in reality this should be larger for the edges. A more complete treatment considering the flux coordinate is left for future work, though it was shown itn [\onlinecite{Reiman18}]  that the slab geometry shared the qualitative physics with the more realistic geometry.\par
	 Nonetheless, the model retains an important feature of the magnetic island geometry: the closed nature of magnetic surfaces about the island centre. Consequently, in the reduced 1D model,  temperature solutions are required to be even about the centre $x=0$. \par
\begin{figure}
\hspace*{-0.3cm}
\includegraphics[width=0.4\textwidth]{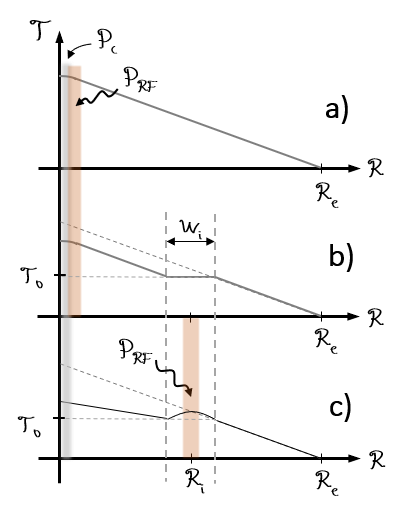}
\caption{Schematic sequence of temperature profiles of the full plasma for: a) No island present b) An island present at $R=R_i$ c) RF energy deposition displaced to within the island. The reddish band represents the region of RF deposition, while the gray one corresponds to other heating sources.}\label{fig::figBry}
\end{figure}

Proceed now to linearise Eq.~(\ref{eqn::dif2}). Assume that the temperature changes ($\widetilde{T}$) in the island due to directing RF heating ($P$) to it are small; ie. $\epsilon=\widetilde{T}/T_0\ll 1$, where $T_0$ is the equilibrium temperature. Seeking precision in this definition, $T_0$ is defined to be the island temperature when the RF power is not aimed directly at the island, but is rather part of the total power budget that heats the centre of the tokamak (see Fig.~\ref{fig::figBry}). Because $T_0$ is constant over the island (as is the density), any island inhomogeneity that develops will be at least $O(\epsilon)$; and with this ordering we drop derivatives with respect to $\kappa_\perp$ in Eq.~(\ref{eqn::dif2}), \par 
\begin{equation}
\frac{3}{2}nk_B\partial_t \widetilde{T}-\kappa_\perp\partial^2_x \widetilde{T}=P \label{eqn::difEqn1D}
\end{equation}
The consideration of stiff temperature profiles that could modify $\kappa_\perp$ non-smoothly is left for future work.
\par
	To drop the time dependence of Eq.~(\ref{eqn::difEqn1D}), the time scales of concern should exceed those of energy diffusion ($\kappa_\perp\partial^2_x \widetilde{T}$ term) and the driving times ($P$ term). Consider the former; clearly for the dominance of the diffusion term $ t\gg\tau_D=W_i^2/\chi_\perp$. Here $W_i$ represents the island width and $\chi_\perp\approx\kappa_\perp/nk_B$ is the heat diffusion coefficient. \par
For the latter, $nk_B\partial_t T\sim P$ suggests that the time derivative may be dropped provided the power density of the wave has had enough time to deposit all the needed thermal energy; i.e., $ t\gg \tau_E= nk_B\Delta T/P$, where $\Delta T\sim\epsilon T_0
$ is the characteristic variation of the island temperature.  Then, with $t$ in this regime, and being consistent with all of previous requirements, \par
\begin{equation}
-\kappa_\perp\partial^2_x \widetilde{T}=P \label{eqn::difEqn1DSimp}
\end{equation}
Prior to detailing the form of $P$, the problem should be closed by both defining the spatial domain and setting appropriate boundary conditions. Naturally, one defines the last closed surface of the island including the X-points as boundaries of the domain of $x$, ie. $|x|\leq W_i/2$. The island width, $W_i$, will be kept constant. That means that $t$ must be shorter than the typical island growth $\tau_i=\left(\partial \ln W_i/\partial t\right)^{-1}\approx\tau_\eta$, where $\tau_\eta$ is the global resistive time scale.\cite{Ruther73} Table \ref{tab::timeScales} shows this last requirement is consistent with previous time orderings. Width changes may then be treated adiabatically; ie. the steady state Eq.~(\ref{eqn::difEqn1DSimp}) may be taken to be satisfied at all times as $W_i$ is changed artificially. \par

\begin{table}
\begin{tabular}{cc}
    Time scale & s \\\hline \hline
    $\tau_\mathrm{se}$ & $1\times 10^{-3}$ \\
    $\tau_\perp$ & $1\times 10^{-3}$ \\
    $\tau_D$ & $3\times 10^{-3}$ \\
    $\tau_E$ & $8\times 10^{-3}$\\
    $\tau_\mathrm{eq}$ & $4\times 10^{-1}$\\
    $\tau_i,\tau_\eta$ & $2\times10^2$\\
 \\ 
\end{tabular}
\begin{tabular}{cc}		
    Dimensionless & scales  \\\hline \hline\\
    $f_\kappa$ & $10^{13}$ \\
    $\epsilon$ & $10^{-1}$ \\
\\
\\
\\
\\
\end{tabular}
\caption{Summary of the various scales relevant to obtain the form of Eqs. (\ref{eqn::approx1}) and (\ref{eqn::approx2}) in typical tokamak parameters of $T=10~$keV, $a=1~$m, $R=5~$m, $W_i=0.05~$m, $n=10^{14}~$cm$^{-3}$, $\chi_{e\perp}=0.5~$m$^2$s$^{-1}$, $Z=1$, $q=2$.} 
\label{tab::timeScales}
\end{table}

The boundary condition on temperature encodes the influence of the island on the remaining of the plasma (and vice versa). To specify it, a simplified treatment of the energy dynamics of the rest of the tokamak is done, using a steady state diffusion model like Eq.~(\ref{eqn::difEqn}). It is convenient to apply Gau{\ss}' theorem to magnetic flux surfaces, so that $\int_\psi PdV=-\partial_\psi T\int_{\partial \psi}\kappa_t\hat{n}\cdot\nabla\psi~dS$, where $\psi$ is the magnetic flux coordinate. This shows that the slope of the temperature profile at a particular flux surface is determined by the power deposited inside it. For simplicity, let us associate a spatial 1D coordinate $R$, in the absence of islands monotonic with $\psi$, and take the heat conductivity, $\kappa_t$, to be constant; then, the slope of temperature at some $R_0$ is determined by the power deposited at $R<R_0$. With this in mind, let there be some heating in the tokamak centre: $P_c$ (fusion power, Ohmic heating, etc.) and $P_\mathrm{RF}$ (RF heating). The temperature profile is then determined by the heat flux and the fixed plasma edge temperature (see Fig.~\ref{fig::figBry}a). \par

Now, let there be an island of size $W_i$ at a distance $R_i$ from the core over which the temperature profile is flat (see Fig.~\ref{fig::figBry}b). Because heat sources have not changed, the temperature slope remains unchanged elsewhere. Let then $P_\mathrm{RF}$ be redirected to the island (ie. the case of interest). For those magnetic flux surfaces at $R>R_i+W_i/2$, the enclosed total power does not change, and thus the slope of $T$ should neither (see Fig. \ref{fig::figBry}c). Given that the tokamak plasma edge temperature is fixed, the temperature of at  the edge of the island, $T_0$, remains unchanged. 
The boundary condition for our island temperature may then be taken to be $\widetilde{T}(x=\pm W_i/2)=0$.\par

It is now the turn of specifying $P$ in Eq.~(\ref{eqn::difEqn1DSimp}) to represent the energy deposition from RF waves. Adopting a geometrical optics (GO) description of the wave envelope\cite{Tracy}, the evolution of the energy of the wave may be written as
\begin{equation}
d_t\Bar{V}=[-(\nabla\cdot v_g)+\frac{\omega_t}{\omega}+2\gamma]\Bar{V}\approx 2\gamma \Bar{V}, \label{eqn::waveEnrgy}
\end{equation}
where $\Bar{V}$ represents the wave energy density, $v_g$ is the group velocity of the wave, $\omega_t$ represents the time derivative of frequency due to a time dependent medium, $d_t$ represents the total time derivative following a wave along a ray and $\gamma$ represents collisionless damping rate. Assuming the medium to be stationary in the wave damping time scale ($\omega_t/\omega\ll \gamma$) and the spatial inhomogeneity to be much smaller than the variation resultant from the damping ($\nabla\cdot v_g\ll\gamma$), the last approximated equality follows. This condition is not difficult to satisfy considering only small variations are created within the island. \par
Expressing Eq.~(\ref{eqn::waveEnrgy}) in terms of $x$, the distance along the ray $d_tx=v_g$, 
\begin{equation}
d_x\Bar{V}=2\frac{\gamma}{v_g}\Bar{V}, \label{eqn::waveEnerSimp}
\end{equation}
which has the form of damped propagation.
The factor $\gamma$ may be obtained under the assumption of a Maxwellian magnetised background\cite{Stix,karney81}, one may show that for EC and LH, $\gamma\propto \exp(-\chi^2)$ where $\chi=(\omega-n\Omega_e)/kv_{Te}=v_\parallel/v_{Te}$ where $v_\parallel$ is the phase velocity of the wave, $v_{Te}$ is the electron thermal speed, $\Omega_e$ is the electron cyclotron frequency, and $n=0$ corresponds to LH waves and $n=-1$ to EC waves. The power deposition both for electron cyclotron and lower hybrid waves occurs on the tail of the Maxwellian velocity distribution, with damping exponentially small in the lowest resonant velocity.  \par
It is this exponential factor which makes deposition highly sensitive to variations in temperature. Indeed, considering the phase velocity of the wave to remain constant over the extent of the island,
 $$\gamma\propto e^{-(v_\parallel/v_{Te_0})^2}\exp\left(\frac{v_\parallel^2\widetilde{T}}{v_{Te,0}^2T_0}\right)\rightarrow \frac{2\gamma}{v_g}=-\alpha e^u.$$
 It is convenient here to define the dimensionless variable $u\equiv v_\parallel^2\widetilde{T}/v_{Te,0}^2T_0=w^2\widetilde{T}/T_0$, where $w=v_\parallel/v_{T_e,0}$. 
The location of the wave damping within the whole plasma is generally dependent on $T$, $B$, $\omega$ and $k$. This location can be determined using ray tracing\cite{Prater08,bonoli1986simulation}. Here the picture is simplified by artificially restricting the damping to a particular defined region within the island , while keeping $v_\parallel$ constant, and that way allowing for the $\widetilde{T}$ expansion. A fully self consistent, full GO analysis is left for future work. 
\par
Let us express Eq.~(\ref{eqn::waveEnerSimp}) as
\begin{equation}
d_x\Bar{V}=-\alpha(x)e^u\Bar{V} \label{eqn::damp}
\end{equation}
where the damping strength $\alpha(x)\equiv2\alpha_0f(x)/W_i$. The factor $\alpha_0$ represents the strength of the damping, but it is also defined in a dimensionless way to include the island width $W_i$.  For example, for EC waves\cite{karney81} $\alpha_0\approx W_i\sqrt{\pi}\omega_{pe}^2 \exp(-w^2)/2ckv_{Te}$. It is helpful to introduce a more physically motivated interpretation of $\alpha_0$.  If a linear limit is taken of Eq.~(\ref{eqn::damp}), the power deposition profile takes the form $|V'|\propto \exp\left(-2\alpha_0x/W_i\right)$; i.e., $\alpha_0$ is the ratio of the island half-width to the characteristic deposition width. Note that the deposition has an exponential shape, and not its usual Gaussian form generally considered for electron-cyclotron waves\cite{Prater08}; however, both schemes are peaked and of finite width, and ultimately quite similar.  \par
Now, going back to the original question: how is $P$ related to this wave energy $\Bar{V}$? From the damping of the wave along a ray, it is easily seen that the volumetric power deposition at a given point is given by $P=-d_t\Bar{V}\approx -v_gd_x\Bar{V}\equiv -v_g\Bar{V}'(x)$. But, because in this particular geometry the points $\pm x$ are linked together (recall this is true due to them belonging to the same flux surface) these points share the total deposition at $x$ and $-x$. All things considered,
\begin{subequations}
\begin{align}
-\kappa_\perp\partial^2_x \widetilde{T}(x)=-v_g\frac{\Bar{V}'(x)+\Bar{V}'(-x)}{2} \label{eqn::approx1}\\
\Bar{V}'(x)=-\alpha(x)e^u\Bar{V}(x)\label{eqn::approx2}
\end{align}
\end{subequations}

These equations may be non-dimensionalised, reducing them to 
\begin{subequations}
\begin{align}
V'(\widetilde{x})&=-f(\widetilde{x})e^uV(\widetilde{x})  \label{eqn::WaveEnergy} \\
u''&=\frac{V'(\widetilde{x})+V'(-\widetilde{x})}{2} \label{eqn::diffus}
\end{align}
\end{subequations}
where the new $V(\widetilde{x})=\bar{V}(x)W_iv_gv_p^2/2\alpha_0\kappa_\perp T_0v_T^2=\bar{V}(x)W_i\Upsilon^2/\alpha_0$ and $\widetilde{x}=2\alpha_0x/W_i$. Note that the edges are now at $x=\pm\alpha_0$. \par
In order to complete the setting of the problem, an initial value must be taken for Eq.~(\ref{eqn::WaveEnergy}). Let $V_X\equiv V(-\alpha_0)\equiv V_0/\alpha_0$, where $V_0$ is a constant representing some wave energy density input. For interpreting solutions, it is important to bear in mind that $V_X$ is independent of island width, but will however scale as $1/\alpha_0$ with the deposition strength. \par
For clarity in the following Sections III and IV, one may refer to Appendix B as a quick reference for the variables employed.

\section{Hysteresis phenomena}

To investigate the effect of the non-linear wave deposition, consider the tractable basic problem of wave damping occurring everywhere within the island. This case, represented by $f(\widetilde{x})=1$, allows for an analytic solution of Eqs.~(\ref{eqn::WaveEnergy}) and (\ref{eqn::diffus}) (see Appendix A for a detailed derivation). Implementing the appropriate boundary and initial conditions,
\begin{equation}
u(\widetilde{x})=2\log\gamma -\log\left[\sqrt{(\lambda+1)^2-\gamma^2}\cosh\gamma\widetilde{x}+(\lambda+1)\right] \label{eqn::tempProf}
\end{equation}
where the paramters $\lambda$ and $\gamma$ are determined by,
\begin{subequations}
\begin{align}
\left[\gamma^2-(1+\lambda)\right]^2=\cosh ^2(\alpha_0\gamma) \left[(\lambda+1)^2-\gamma^2\right] \label{eqn::param1} \\
\gamma^2=(2\lambda+1)+(V_X-\lambda)^2 \label{eqn::param2}
\end{align}
\end{subequations}
 \par
The integration constant is $\lambda=(V_X+V_f)/2$, where $V_f$ is the energy density when exiting the island. Eqs. (\ref{eqn::param1}-\ref{eqn::param2}) solve $\lambda$ implicitly, which ultimately determines the temperature of the island as a function of $\alpha_0$ and $V_X$ from Eq.~(\ref{eqn::tempProf}). \par

\begin{figure}
\hspace*{-0.3cm}
\includegraphics[width=0.5\textwidth]{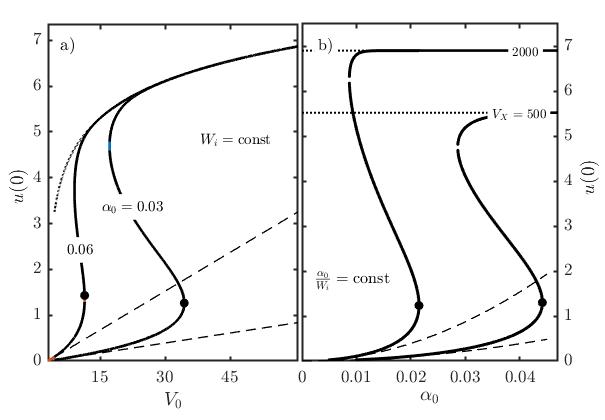}
\caption{a) Island temperature at fixed deposition strength for different island widths, as a function of absorbed RF power, showing the appearance of the bifurcation point (shown bigger points). Dashed lines correspond to the linear limit of the solution; the dotted line to the asymptotic form of the solution for almost complete power deposition. 
b) Central island temperature with varying island width for constant incident wave energy densities $V_X$. The broken lines represent the solution to the linear problem, while the dotted ones the asymptotic limit solution for complete energy deposition.}
\label{fig::fig2}
\end{figure}

The dependence of perturbed central island temperature on these two parameters will be represented as contour curves (see for example Fig.~\ref{fig::fig2}). Two main representations are of particular physical interest. First, contours of constant $\alpha_0$ (ie. fixed deposition strength and island width) in the $u(0)-V_0$ plane. These contours show the effects of the wave power on temperature (see Fig.~\ref{fig::fig2}a). The second interesting picture is related to how the heating of the island evolves as its width or the wave profile width changes. This is captured by curves of constant $V_X$ or $V_0$ at fixed deposition strength in the $u(0)-\alpha_0$  plane (see Fig.~\ref{fig::fig2}b for an example).  \par

There are a number of general features in the solutions to Eqs.~(\ref{eqn::tempProf})-(\ref{eqn::param2}) worth highlighting. The first of those is the existence of bifurcation points. As previously observed\cite{Reiman18}, for sufficiently broad depositions (small $\alpha_0$), saddle-node bifurcations appear, at which the two lower temperature solutions disappear. Such points are marked in Fig.~\ref{fig::fig2}a. The appearance of these points may be linked to the action of a self-focusing mechanism affecting RF waves. Schematically, below the bifurcation, significant energy leakage takes place. As the bifurcation is approached, island temperature perturbations become larger, while the power deposited by the wave increases accordingly. This positive feedback eventually extracts all RF energy effectively, reaching a higher temperature steady state and thus jumping into an upper branch of the solution. \par

This hot stable solution may be seen in Fig.~\ref{fig::fig2}, along with the asymptotic form of the solution as $V_f\rightarrow 0$ (dotted line). The proximity of the two solutions demonstrates that the upper branch indeed corresponds to nearly complete deposition of the wave energy in the island. Such a solution branch is also, immediately after the bifurcation point, significantly larger than the linear prediction (see broken lines). \par

A consequence of the solution structure obtained is the hysteresis behaviour of island heating. To illustrate such a process, 
 take as a starting point the system to be in equilibrium at the lower temperature branch in Fig.~\ref{fig::fig2}a, and increase the absorbed energy of the incoming wave ($V_0$) gradually. As a result, the temperature of the island will grow until the bifurcation point is reached. Once at this point, and driven by the self-focusing feedback, the temperature of the island will rapidly increase  towards the upper branch, which is the only stable solution at high absorbed power.\par
 The hot island exhibits, at this point, a large temperature difference between the centre and the separatrix. A priori, this would help to absorb RF power closer to the O-point, and thus also drive thte central current\cite{fisch87}. Here we ignore current associated with the DC electric field, which could be due to Spitzer conductivity or due to the hot electron conductivity\cite{fisch85a,karney_fisch_jobes}. These currents are less important than the directly driven RF current.\cite{Reiman18} With such presumed centred current drive, the island would tend to shrink and stabilise, as governed by the Rutherford equation\cite{Ruther73}. This size reduction corresponds to a leftwards displacement towards smaller values of $\alpha_0$ in Fig.~\ref{fig::fig2}b. In such a case, and if the energy available to the island is maintained, the reduction does not imply a return back to the original low temperature, but instead remains in the more effective current driving upper branch for some time. Similarly, once in the upper branch, driving power requirements are relaxed, and lower $V_0$ would still keep the plasma hot. This constitutes the hysteresis effect. A more careful discussion on the usefulness, accessibility and consistency of this sketched simplified picture for particular deposition schemes is the concern of following sections. \par

 \par

\section{Effects of non-linear feedback}

\subsection{Typical parameters}
	Before proceeding further, a brief estimate and collection of typical values for both $V_0$ and $\alpha_0$ is presented. We emphasise that our slab model provides a physical qualitatively correct picture of the problem, but only a rough guide into the quantitative behaviour of the more realistic geometry, as previous calculations suggest\cite{Reiman18}. Other simplified features, such as the exponential form of the linear deposition profile, are also different when compared to actual experiments\cite{Prater08}, but do however share the fundamental characteristics. Thus, the linear case will be taken as reference in guiding conclusions in the following sections, as well as orientative comparison standard to existing experimental parameters.    \par
	Focus first on the values for the parameter $\alpha_0$. Two different routes are taken at this point. One possible method uses, given the definition of $\alpha_0$ as the size of the linear power deposition width, typical deposition widths in tokamak experiments could be used to obtain $\alpha_0\sim 0.5-3$.\cite{Volpe15} It has been recently reported that current drive profiles are in experiment subject to broadening\cite{Brook17} by factors of 2--3 due to effects unaccounted for in ray tracing routines, such as edge density fluctuations. This effective broadening could make typical $\alpha_0$ values even lower, down to $\sim 0.2$.   \par
	Alternatively, one could use the form for $\alpha_0$ given before and obtained in the context  of GO. In the case of ECCD, for instance, using typical approximated hydrogen tokamak values (see caption of Table \ref{tab::timeScales}), with $w^2\sim10$,\cite{karney81} wavenumber\cite{Prater04} $k\sim2\pi/(5~\mathrm{mm})$, density $n\sim10^{20}$~m$^{-3}$ and $W_i\sim10~$cm; $\alpha_0\approx W_i\sqrt{\pi}\omega_{pe}^2 \exp(-w^2)/2ckv_{Te}\sim10^{-1}$. 
\par
	Now consider the wave power density $V_0$. First, we may compare existing literature\cite{Westerhof07} where island temperature variations are computed to our linear toy model. Comparing values of $u(0)$,orientative typical power parameters on the order of $V_0\sim5$ for 20~MW RF power are found. In a more first principle approach, we might use the rescaled definition of $V_0$ introduced before. Using an RF power on the order of $P\sim10$~MW, with a beam of cross section $A\sim1~$m$^2$, with $\chi_\perp\sim1$m$^2$/s, temperature $T_0\sim10$~keV, density $n\sim10^{20}$~m$^{-3}$ and $w^2\sim10$, then $V_0\approx PW_iw^2/2A\chi_\perp nk_BT_0\sim10^1$. \par
	Summarising:
\begin{equation}
	\alpha_0\sim0.1-3 \hspace{1cm}  V_0\sim 10^1
\end{equation}

\subsection{Central deposition}
	The spatial distribution of the RF deposition strongly affects the final temperature of the island, as well as the island mitigation efficiency. In this section, the best case scenario is first analysed; i.e., deposition starting from the island centre. To formally emulate this ideal case, $f(\widetilde{x})=H(\widetilde{x})$, where $H$ is the Heaviside step function. Given this newly introduced asymmetry, well defined parity is lost from the equations and the solution to the equation is only found numerically. \par
\begin{figure}
\hspace*{-0.3cm}
\includegraphics[width=0.5\textwidth]{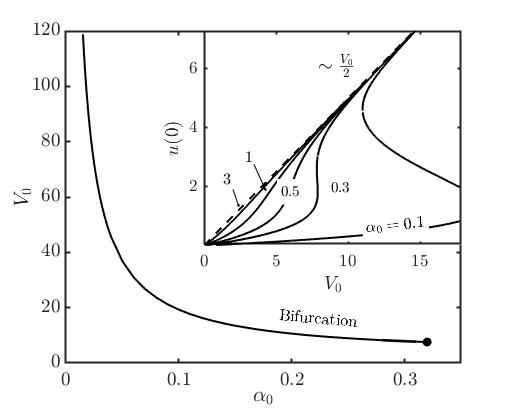}`
\caption{Wave power density value for the bifurcation at a given deposition strength $\alpha_0$, for deposition starting at the island centre. The larger scatter point represents the limiting value of $\alpha_0$ over which no bifurcation exists. The inset shows curves of constant deposition strength in the $u(0)-V_0$ plane. The broken line represents the asymptotic form of the non-linear solution.}
\label{fig::fig3} 
\end{figure}
\begin{figure}
\hspace*{-0.3cm}
\includegraphics[width=0.5\textwidth]{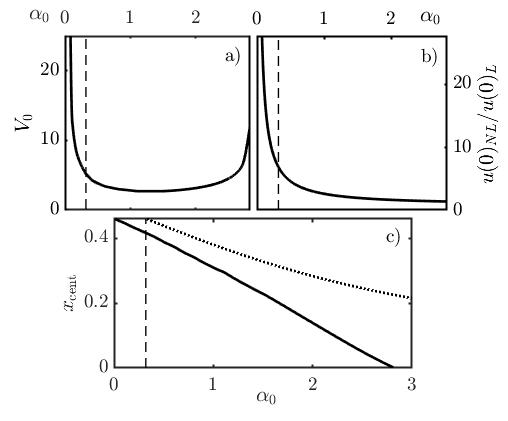}
\caption{a) Wave power density value for 50\% difference between the linear and non-linear solutions for centred wave energy deposition. b) Upper bound to ratio of non-linear to linear island temperature. c) Location of middle point of the wave deposition $x_\mathrm{cent}|V(x_\mathrm{cent}\alpha_0)=(V_X+V_f)/2$ for the points in the curve in a) (continuous line), and middle point for the linear equivalent problem (dotted line).  Solutions for the region $\alpha_0$ to the left of the broken line represent solutions with existing bifurcation points.}
\label{fig::fig4} 
\end{figure}

	Consider first the occurrence of bifurcation points, and in particular, how they depend on $\alpha_0$. To illustrate these points, Figures \ref{fig::fig3} and \ref{fig::fig4} are presented. \par
	In the broad linear deposition limit, with $3\alpha_0\ll1$, there always exists a bifurcation point (see Fig.~\ref{fig::fig3}). The turning point, however, occurs at increasingly larger wave energy densities $V_0$. One may understand this result by referring to the analytic asymptotic form of the solution at low $V_0$. In that limit, the system takes the form of the linear problem, for which $u(0)\sim V_0\left(\alpha_0-1+e^{-\alpha_0}\right)/2\alpha_0$. This shows that the temperature of the island becomes decreasingly responsive as $\alpha_0\rightarrow 0$ (see the decreasing initial slopes of curves in Fig.~\ref{fig::fig3} inset), which is ultimately related to there being a significant wave energy leakage ($V_f=V_Xe^{-\alpha_0}$). \par
	Only for those cases for which the initial energy leakage $V_f$ is significant will a bifurcation occur. The bifurcation is a result of the system being able to access all that previously lost energy when the damping $e^u$ factor becomes significant. This jump will be associated with a narrowing of the deposition and a current that is more efficiently utilised in stabilisation. \par

	As the initial deposition is reduced by increasing $\alpha_0$, $V_f$ in the lower branch decreases, and it eventually becomes too small to sustain a bifurcation. Fig.~\ref{fig::fig3}) shows the boundary value $\alpha_0=0.32$ beyond which no bifurcation occurs. \par
	Where no bifurcation occurs, the temperature of the island only undergoes a smooth transition in temperature between the linear solution and the high temperature asymptote (see Fig.~\ref{fig::fig3}). That limiting form of the nonlinear solution as $V_f\rightarrow0$ is $u(0)\sim V_0/2$ (see dashed line in Fig.~\ref{fig::fig3} inset and Appendix C), which is also the limit as $\alpha_0\rightarrow\infty$ of the linear deposition. That is, the non-linear response serves as a short cut via self-focusing to the linear ideal infinitely narrow deposition.  Thus, for the case of centralized deposition, the nonlinear mechanism always leads to an enhanced temperature increase. 
 \par
	However, the wave power required to obtain a substantial improvement exhibits strong dependence on $\alpha_0$ as shown in Fig.~\ref{fig::fig4}a. The large energy leakage and the small difference between linear and non-linear solutions at low and high $\alpha_0$ respectively leave a most easily accessible (lower $V_0$) central region at values $\alpha_0\sim1-1.5$. To emphasise the second of these limitations Fig.~\ref{fig::fig4}b shows the ratio of the analytic asymptotic forms of the non-linear and linear solutions. Evidently, the differences become marginal (ie. the ratio tends to one) for stronger depositions, which explains why Fig.~\ref{fig::fig4}a diverges at $\alpha_0\sim2.8$. \par
	Finally, we examine the extent to which self focusing narrows the RF deposition profile. From Fig.~\ref{fig::fig4}c the self-focusing effect is apparent, and will undoubtedly improve stabilisation by bringing current drive closer to the O point. This reduction by a factor of $\sim2-3$ in width opens the possibility of previously disregarded regimes of island stabilisation. Note also that for $\alpha_0\approx2.8$, the non-linear deposition profile becomes similar to that of a delta function (within discretisation). \par
	This analysis suggests that the region of interest and current experimental relevance may in some subset of cases (for the broadest depositions) show some hysteresis behaviour, but most will just show significant temperature variation. In addition, as a result of deposition narrowing due to the non-linearity, islands could be stabilised when traditionally predicted not to. This opens the door to experimental verification of the non-linear effect, as well as extension of mitigation schemes.  \par

\subsection{Edge deposition}
	The scenario adopted for analysis before was that of central deposition. This is the ideal case, and so it presumes that one is experimentally capable of aiming perfectly at the centre of the island without depositing any energy before that. But, what would happen if the deposition departs from this idealised case? The worst case scenario is now presented. To that end, we recover the analytic solution from Eqs. (\ref{eqn::tempProf})-(\ref{eqn::param2}) which represents wave deposition from the very edge of the magnetic island. \par
With this analytic result at hand, let us explore first the limit of complete wave deposition: $u(0)\sim \log\left(V_0/2\alpha_0+1\right)$. The linear solution gives $u(0)=V_0e^{-\alpha_0}(\cosh\alpha_0-1)/\alpha_0$. It is remarkable that the non-linear model gives a logarithmic growth of the island temperature as the power input is increased, while the linear case grows linearly. It necessarily follows that some non-linear inhibition mechanism must take part. Indeed, one may relate this to the deposition profile becoming localised ever closer to the edge of the island. \par
 	It is the same self-focusing that narrowed the deposition closer to the centre when central deposition was considered, which displaces deposition towards the island edge (see Fig.~\ref{fig::fig6}). Physically, the RF wave becomes so strongly damped that it runs out of energy very close to the edge. There, the temperature slope is large, and thus as heat gets to the centre of the island, it is quickly lost across the edges, $u(0)$ becoming limited. For typical values $\alpha_0\sim1$ and $V_0\sim5-10$, the peak of deposition $x_\mathrm{peak}\sim-0.7$; ie. power is deposited somewhere between the X- and O-points. \par

\begin{figure}
\hspace*{-0.3cm}
\includegraphics[width=0.5\textwidth]{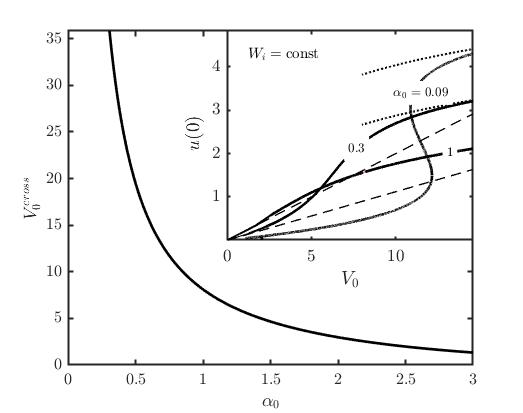}
\caption{Power at which the non-linear and linear solutions give the same central temperature with deposition starting from the edge, as a function of the deposition strength. The region to the left of the curve represent the case for larger non-linear solution. The inset shows examples of constant deposition curves as a function of power. The broken lines represent the linear solution, while the dotted curves show the logarithmic asymptotic behaviour of the non-linear one. }
\label{fig::fig5} 
\end{figure}

\begin{figure}
\hspace*{-0.3cm}
\includegraphics[width=0.5\textwidth]{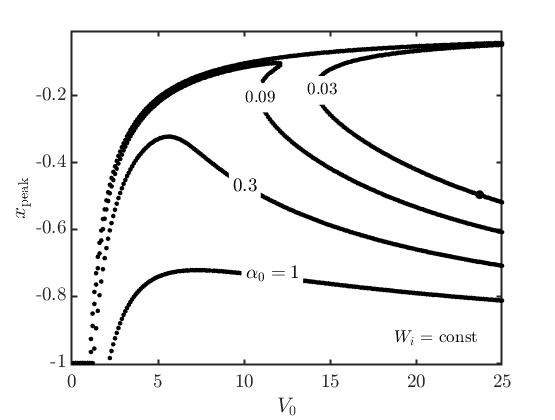}
\caption{Location of the power deposition peak as a function of power for various deposition strengths (complementary to Fig.~\ref{fig::fig5}). A value of $x_\mathrm{peak}=0$ corresponds to a centred deposition, while a value of $-1$ represents the edge.}
\label{fig::fig6} 
\end{figure}
	As a result of this detrimental displacement, only over a limited regime will the non-linear solution be hotter than the linear one. As shown in Fig.~\ref{fig::fig5}, this interval is larger for the broader depositions, but tends to disappear as $\alpha_0\rightarrow\infty$. When the deposition is broader, the non-linear focusing increases the amount of deposited power significantly, a benefitial addition that outweights the inhibiting displacement for a more extended range of powers.  \par
	In addition to temperature, this deposition shift will also bring the driven current closer to the X-point. This displacement can be catastrophic when trying to mitigate the growth of magnetic islands. The proximity of the deposition to the edge may be seen in Fig.~\ref{fig::fig6}. The plot shows that indeed, for large $V_0$ values, the non-linear self-focusing brings the deposition ever closer to the edge. This is the result of the large damping $e^u$ which drains the incoming wave faster than the linearly increasing $V_0$. Nevertheless, there is an initial region in which the focusing is starting to affect the system, and clearly becomes beneficial in terms of drawing power towards the centre. \par

\begin{figure}
\hspace*{-0.3cm}
\includegraphics[width=0.5\textwidth]{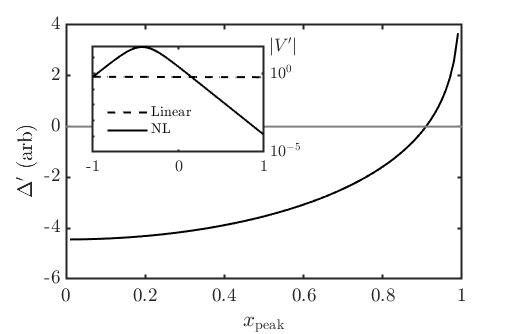}
\caption{Scaled contribution to $\Delta'$ in the Rutherford equation due to current drive at different positions within the island. Current drive is stabilising for $\Delta'<0$, ie. when driving is roughly within 90\% of the island extent. The inset shows the deposition profiles for the linear and non-linear solutions corresponding to the point shown in Fig.~\ref{fig::fig6}, represented in log scale (the linear deposition is almost uniform, but largest at the X point). }
\label{fig::fig7} 
\end{figure}

	To further explore the sensitivity of stability on $x_\mathrm{peak}$, the procedure in [\onlinecite{Lazzari}] is followed. The relative Fourier weighting to $\Delta '$ in the Rutherford equation due to driving current at a particular flux coordinate, $\psi$ ($\Omega$ in the reference), may be estimated taking the 1D spatial variable $x$ in our model to match the spatial $x$ in [\onlinecite{Lazzari}], looking at $\xi=0$ (see Fig.~\ref{fig::fig7}). The calculation shows that within approximately 90\% of the island the drive is stabilising. Therefore, looking back at Fig.~\ref{fig::fig6}. The current drive will still be central enough to be stabilising for a significant fraction of the cases, even when the non-linear solution is colder than the linear one. No definitive conclusion may however be drawn on the precise fraction of the island that is truly stabilising, as the treatment of the island geometry in [\onlinecite{Lazzari}] is different from that of our model. A fully consistent treatment is left for future work.  \par
	We have shown that an initially broad RF profile may then be used to stabilise islands, so long as the input power remains below some upper bound. This result is promising and an idea to further explore. \par

\par
More generally, if deposition was to start midway between the X- and O-points, then deposition would be driven and narrowed towards that location instead. The trends will then fit between the extreme centre and edge cases shown explicitly, with less constraining requirements as the centre is reached. In this context, RF overshooting scenarios will never suffer from the inhibition that takes place when undershooting with increasing $V_0$.  \par
Yet another possibility to circumvent inhibition might be to look for means to amplify the wave power within the island, such as through an $\alpha$-channeling effect\cite{fisch1992interaction}. This could give a power not peaked at the island periphery. Unfortunately, this volumetric amplification does not occur for electron cyclotron waves, but might be exploited when using lower hybrid waves\cite{ochs_2015a}. Despite lacking this possible enhancement, EC waves do however have the benefit of having a $B$-dependent resonance that allows for deposition starting at a particular point in space.

\section{Conclusion}
The possibility of hysteresis involving the heating of magnetic islands with RF waves is shown for a symmetrised, 1D slab model. The wave power deposition self focuses mediated by island temperature, leading to higher temperatures. Past a bifurcation point, the island may remain in this high temperature solution even as it shrinks or power is reduced. Exploiting the hysteresis effect could thus provide an easier and improved way to eliminate magnetic islands. \par
In typical parameter regimes of current experiments, though, bifurcation is likely to occur only for the broadest profiles. Yet, centred deposition scenarios show that important temperature increments (on the order of $\sim20\%$) occur. These differences are most significant in broader deposition schemes, where the self-focusing mechanism of the non-linear model is most different from the linear model. Alongside these thermal variations, the non-linear narrowing of the RF deposition profile by a factor of up to $\sim2-3$ for typical values will improve the utility of current drive for purposes of stabilisation.\par
Broad schemes that deviate from the centre still lead to not dissimilar self-narrowing and stabilisation under certain circumstances. These circumstances involve the form of EC power density profiles, and must be considered in designing deposition scenarios. In particular, for deposition profiles that peak before the O-point, there is a threshold power density above which a self-inhibition mechanism is encountered; beyond the O-point, this threshold does not exist. This opens the door to exploring previously disregarded broad RF deposition stabilisation schemes.  \par

\begin{acknowledgments}
Thanks to Suying Jin for fruitful discussions and help. \par
This work was supported by US DOE DE-AC02-09CH11466 and DE-SC0016072.

\end{acknowledgments}

\appendix

\section{Analytic solution to constant $\alpha$}
Consider the coupled set of equations,
\begin{subequations}
\begin{align}
V'(\widetilde{x})&=-e^uV(\widetilde{x})  \label{eqn::WaveEnergyA} \\
u''&=\frac{V'(\widetilde{x})+V'(-\widetilde{x})}{2} \label{eqn::diffusA}
\end{align}
\end{subequations}
Define the following symmetric and antisymmetric parts of the wave energy density,
\begin{align}
S=\frac{V(\widetilde{x})+V(-\widetilde{x})}{2} \label{eqn::sym}\\
A=\frac{V(\widetilde{x})-V(-\widetilde{x})}{2} \label{eqn::asym}
\end{align}
Given these, Eq.~(\ref{eqn::diffusA}) may be cast in the form,
\begin{equation}
u''=\left[\frac{V(\widetilde{x})-V(-\widetilde{x})}{2}\right]'= A'
\label{eqn::diffHeatRe}
\end{equation}
Substituting definitions (\ref{eqn::sym}) and (\ref{eqn::asym}) into Eq.~(\ref{eqn::WaveEnergyA}),
\begin{equation*}
(A+S)'=-e^u (A+S)
\end{equation*}
and realising that the spatial derivative $\frac{d}{dx}$ is an odd operator while $u$ is an even function, the equation may be separated into its symmetric and asymmetric parts,
\begin{align}
S'=-e^uA \label{eqn::resHeati}\\
A'=-e^uS \label{eqn::resHeatii}
\end{align}
which with Eq.~(\ref{eqn::diffHeatRe}) form a set of three coupled non-linear differential equations. The set avoids the explicit appearance of $-x$ on the expense of an additional equation. \par
Because of even parity $u'(\widetilde{x}=0)=0$, so Eq.~(\ref{eqn::diffHeatRe}) may be integrated
\begin{equation}
u'=A
\label{eqn::simpU'}
\end{equation}

\begin{table}
\begin{tabular}{c|c|c}
    Var & $\widetilde{x}$ limit & Var limit\\\hline \hline
    $u$ & $-\alpha_0$ & 0 \\ \hline
    $u'$ & $0$ & $0$ \\ \hline
    $V$ & $-\alpha_0$ & $V_X$ \\ \hline
    $V'$ & $-\alpha_0$ & $-V_X$\\\hline
\end{tabular}
\caption{Initial conditions that specify the problem that is to be solved.} 
\label{table::limit} 
\end{table} 
Substituting (\ref{eqn::simpU'}) and (\ref{eqn::diffHeatRe}) into equations (\ref{eqn::resHeati}) and (\ref{eqn::resHeatii}), putting together and integrating once
\begin{equation}
    u''=e^u[e^u-(1+\lambda)] \label{eq::eqnMain}
\end{equation}
where $\lambda$ is an integration constant that must satisfy the initial conditions specified. From Eq.~(\ref{eqn::diffHeatRe}) it follows that $\lambda=(V_X+V_f)/2$, which limits $V_X/2<\lambda<V_X$. 
\par
To solve Eq.~(\ref{eq::eqnMain}), multiply both sides of the equation by $u'$ and integrate with respect to $\widetilde{x}$. Thus,
\begin{equation}
    u'^2=e^u\left[e^u-2(\lambda+1)\right]+\mathbf{C} \label{eqn::1stOrd}
\end{equation}
where $\mathbf{C}$ is another integration constant to be determined later on.\par
Let the substitution $u\equiv-\log z$ be implemented in Equation (\ref{eqn::1stOrd}), 
\begin{equation}
    z'^2=1-2(\lambda+1)z+\mathbf{C}z^2 \label{eqn::zEqn}
\end{equation}
At his point, one may try a symmetric solution of the form $z=A\left(e^{\gamma \widetilde{x}}+e^{-\gamma \widetilde{x}}\right)+B$, where $A$, $B$ and $\gamma$ are to be reduced to a single integration constant. If this is possible, uniqueness guarantees this to be the general solution. After some manipulation,
\begin{align}
         \mathbf{C}=\gamma^2 , ~B=\frac{\lambda+1}{\gamma^2}, ~A^2=\frac{(\lambda+1)^2-\gamma^2}{4\gamma^4} \label{eqn::relCBA}
\end{align}
which indeed leaves a single degree of freedom, $\gamma$, as expected for a first order ODE. It is then time to implement boundary conditions to determine $\lambda$ and $\gamma$. From $u=0$ at $\widetilde{x}=-\alpha_0$, $z(\widetilde{x}=-\alpha_0)=1$ is evaluated, and eliminating $A$, 
\begin{equation}
    \left[\gamma^2-(1+\lambda)\right]^2=\cosh ^2\alpha_0\gamma \left[(\lambda+1)^2-\gamma^2\right] \label{eq::Lgrel}
\end{equation}
Having eliminated $A$, the solution for $u$,
\begin{equation}
    u(\widetilde{x})=2\log\gamma -\log\left[\sqrt{(\lambda+1)^2-\gamma^2}\cosh\gamma\widetilde{x}+(\lambda+1)\right]
\end{equation}
The additional boundary or initial condition may be imposed requiring $u'(-\alpha_0)=A(-\alpha_0)=(V_X-V_f)/2=V_X-\lambda$ from Eq.~(\ref{eq::eqnMain}), which will introduce explicitly the physically relevant parameter $V_X$. Then,
\begin{equation}
    \gamma^2=(2\lambda+1)+(V_X-\lambda)^2 \label{eq::V0eq}
\end{equation}
where $\gamma^2>1+2\lambda$. \par


\section{Summary of variables and abbreviations}
A collection of the variables used in Sections III and IV is presented as a reference in Table \ref{table::descrVar}. A brief description is also provided where relevant.

\begin{table}
\begin{tabular}{cc}
    $\widetilde{x}$ & Dimensionless position \\\hline
 $u$ & Dimensionless temperature variations\\
    $T_0$ & Background temperature \\
 $u(0)$ & $u$ at island centre\\\hline
    $V$ & Dimensionless wave energy density \\
 $V_X$ & $V$ at left X-point \\
    $V_0$ & Scaled $V_X$: $V_0=\alpha_0 V_X$  \\
 $V_f$ & $V$ leaving island\\\hline
$\alpha_0$ & Half island to deposition width\\
 & (also island edges) \\
$\alpha_0/W_i$ & Deposition strength \\
$W_i$ & Island width\\\hline
$x_\mathrm{cent}$ & Deposition mid-point location \\
$w$ & Ratio phase to thermal speeds \\
$\omega$ & RF frequency \\
\end{tabular}
\caption{Reference description of variables in Sections III and IV} 
\label{table::descrVar} 
\end{table} 

\section{Asymptotic limit non-linear central deposition}

Let us consider the limiting case for complete RF power deposition in island stabilisation starting deposition from the centre of the magnetic island. \par
Begin with,\par
\begin{equation}
u''=\frac{V'}{2}\rightarrow u'=\frac{V}{2}-\frac{V_X}{2}
\end{equation}
 Now take the wave energy to be damped quickly, so that $V\approx0$ for $x>0$. In that case, and as the edge is located at $x=\alpha_0$:
\begin{equation}
u(0)\sim\alpha_0\frac{V_X}{2}
\end{equation}

\bibliography{aipsamp}

\end{document}